# A Regularized Boundary Element Formulation for Contactless SAR Evaluations within Homogeneous and Inhomogeneous Head Phantoms

Rajendra Mitharwal and Francesco P. Andriulli

*Computational Electromagnetics Research Laboratory, Microwave Department, Télécom Bretagne, Brest, France*

**Abstract**

This work presents a Boundary Element Method (BEM) formulation for contactless electromagnetic field assessments. The new scheme is based on a regularized BEM approach that requires the use of electric measurements only. The regularization is obtained by leveraging on an extension of Calderon techniques to rectangular systems leading to well-conditioned problems independent of the discretization density. This enables the use of highly discretized Huygens surfaces that can be consequently placed very near to the radiating source. In addition, the new regularized scheme is hybridized with both surfacic homogeneous and volumetric inhomogeneous forward BEM solvers accelerated with fast matrix-vector multiplication schemes. This allows for rapid and effective dosimetric assessments and permits the use of inhomogeneous and realistic head phantoms. Numerical results corroborate the theory and confirms the practical effectiveness of all newly proposed formulations.

## 1. Introduction

Cellular phones, laptops, bluetooth/wireless hotspots, broadcasting systems are all devices that emit mid-to-high doses of electromagnetic radiation that penetrates materials and biological tissues in their vicinities. Recommendations from various institutions in Europe and United States, dictate strict limits on the amount of electromagnetic radiation that can be tolerated within tissues and anatomical parts surrounding a radiating source. The guidelines from the International Commission on Non-Ionizing Radiation Protection (ICNIRP), that are the legal standard in Europe [1, 2], are dictating restrictions in terms of electric field induced in tissues (frequency range up to 100kHz [3, 4]), specific absorption rate (SAR, frequency range 100 kHz- 10 GHz [5]), and incident power density (frequency range 10 GHz- 300 GHz [5]). The early validation against these lim-its is a crucial step in every industrial process involving radiating elements in the design.

Common technologies and standards that industry uses to assess the electromagnetic exposure and field levels are largely based on internal probes [6] and phantoms [6, 7, 8]. Phantoms are suitably designed dielectric structures obtained by filling with a dissipative liquid a container mimicking the shape of different anatomical parts whose electromagnetic exposition needs to be studied [9, 10]. The measuring probes penetrate the phantom and effectuate measures of the electric and/or magnetic field [11, 12]. Unfortunately however, such an invasive measurement procedure presents several drawbacks. It is costly, since it necessitates ad-hoc mechanical set-ups. It does not guarantee a perturbation-free measurement since an internal probing can perturb the value of the measured fields [13, 14]. Finally, it has the need of repeatedly penetrating the dielectric phantom, something that prevents the use of solid dielectric materials and, as a consequence, liquid filled phantoms provide homogeneous and isotropic dielectric profiles only. This is a poor and unrealistic modeling of biological tissues whose dielectric/resistive properties are often both inhomogeneous and anisotropic [15, 16, 17, 18].

A first class of partial solutions to some of the above mentioned issues relies on techniques for determining optimal sets of measurement samples. This for the purpose of reducing the overall number of measurements that are necessary for a full characterization of the radiation exposure [19, 20, 21, 22]. Although these techniques decrease the complexity of a standard, phantom-based, electromagnetic exposure analysis, they are still very complicated to implement, and they leave the modeling limitations of the phantom/probe approach, lamentably, unaltered. A second and more recent class of enhanced dosimetry assessment techniques relies on the use of computational imaging tools to complement the raw measurements of the electromagnetic field. Under this category fall several schemes that adopt finite element based discretizations of models of human tissues and, given an external measurements of the electromagnetic field, solve the electromagnetic problem by using FDTD [23, 24, 25], FEM [26, 27, 28], and related methods [29, 30, 31, 31, 32, 33]. Although effective and easily available from commercial simulation toolboxes, these schemes often rely on a good knowledge of radiation sources, something that is often unavailable in dosimetry assessment.

An effective remedy to these issues is proposed by methods relying on the use of the Huygens principle [34, 35], where the (potentially unknown) source is replaced by a surfacic distribution of equivalent sources that are determined together with the field values necessary for the dosimetric assessment. These strategies however, often require the use of densely discretized Huygens surfaces and, when differential equation based methods are the leading modeling formulation, CFL conditions and low-frequency issues (see [36] and references therein) may render the approach computationally expensive. A good alternative could be the use of Huygens principle formulations based on Boundary Element Methods (BEMs). These approaches discretize only material boundaries and are not subject to CFL constraints so that an increase of discretization density in some parts of the Huygens screen would not result in increase in other parts of the simulation volume. Very promising formulations following these strategies have been presented in [37, 38, 39]. These schemes encompass the model of a dosimetric phantom with an equivalent surface and establish a suitable integral relationship linking equivalent sources with magnetic and electric field measurements. Such a strategy falls in the more general category of inverse source approaches that have been been studied extensively in the electromagnetic characterization of radiating sources [40, 41, 42, 43, 44, 45]. The problem of characterizing the sources, especially in the presence of severe ill-posedness has also been impacted by more general techniques in inverse scattering and imaging where the regularization techniques of inverse problems related to imaging have been adapted to microwave imaging for human body tissues and their anomalies[46, 47, 48, 49, 50, 51, 52, 53].

Integral equation techniques however, are not panacea; BEM methods in fact suffer from a major drawback viz. they give rise to dense interaction matrices, resulting in high computational costs when realistic modeling are called for. Moreover, classical formulations rely often on Diriclet-to-Neumann mappings (linking magnetic to electric field quantities and vice-versa) that are well-known to be ill-posed and unbounded operators [54]. For this reason although the use of dense equivalent surfaces does not impact a volume discretization (as in CFL prone methods) it still gives rise to ill-conditioned system as a function of the discretization density that can result in high computational costs for the solution and in numerical instabilities in real case scenarios.

This work focuses in addressing these drawbacks and its contribution is threefold: (i) It will propose a BEM based formulation that requires the use of electric measurements only. This is done at the cost of solving an additional integral problem with respect to the works in [37], but it has the practical advantage of avoiding magnetic field measurements and the theoretical advantage of providing more freedom in the choice of the mapping operators and in the integral formulation modeling the dosimetric phantom. (ii) This additional freedom will be exploited to use a Huygens formulation that can be regularized. For this purpose this work will propose a Calderon-based strategy for rectangular matrices that will provide well-conditioned systems independent on the discretization density of the Huygens screen. (iii) Finally, we will hybridize our formulation with both surfacic homogeneous and volumetric inhomogeneous forward BEM solvers accelerated with fast matrix-vector multiplication schemes. This allows for rapid and effective dosimetric assessments and it permits the use of inhomogeneous and realistic head phantoms.

This paper is organized as follows: Section 2 presents background material and sets the notation. Section 3 presents the Huygens formulation we are adopting here. Section 4 presents a Calderon regularization for rectangular matrices. Section 5 presents numerical results that confirms the practical effectiveness of the new approaches. Section 6 presents our conclusions and avenues for future work.

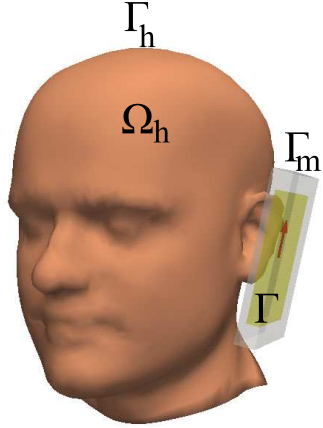

Figure 1: Volume and surface definitions.

## 2. Background and Notation

Consider a volume $\Omega_h$ (please refer to Fig. 1) characterized by a (potentially lossy and inhomogeneous) dielectric permittivity $\epsilon_h(\boldsymbol{r})$ and residing in a free-space of dielectric permittivity and magnetic permeability $\epsilon$ and $\mu$, respectively. The surface of $\Omega_h$ is denoted as $\Gamma_h = \partial \Omega_h$. Consider a (potentially unknown) source radiating the electromagnetic field $\left(\boldsymbol{E}^i(\boldsymbol{r}), \boldsymbol{H}^i(\boldsymbol{r})\right)$; the source resides in the interior of the closed surface $\Gamma$ (an equivalent, Huygens, surface). The total electric field $\boldsymbol{E}(\boldsymbol{r})$ is assumed to be known on a surface $\Gamma_m$ as a result of a measurement procedure. In our application scenario, $\Omega_h$ will model a (potentially lossy and inhomogeneous) head phantom while the source, entirely included in the closed surface $\Gamma$, will model a mobile phone radiator (the model of which is potentially unknown).

On the equivalent surface $\Gamma$ we can define the following surface operators

$$\mathcal{S}_\Gamma^k(\boldsymbol{f}(\boldsymbol{r})) = ik \int_\Gamma \frac{e^{ik|\boldsymbol{r}-\boldsymbol{r'}|}}{4\pi|\boldsymbol{r}-\boldsymbol{r'}|} \boldsymbol{f}(\boldsymbol{r'})d\Gamma - \frac{1}{ik}\nabla \int_\Gamma \frac{e^{ik|\boldsymbol{r}-\boldsymbol{r'}|}}{4\pi|\boldsymbol{r}-\boldsymbol{r'}|} \nabla' \cdot \boldsymbol{f}(\boldsymbol{r'})d\Gamma \quad (1)$$

$$\mathcal{D}_\Gamma^k(\boldsymbol{f}(\boldsymbol{r})) = -\int_\Gamma \nabla \frac{e^{ik|\boldsymbol{r}-\boldsymbol{r'}|}}{4\pi|\boldsymbol{r}-\boldsymbol{r'}|} \times \boldsymbol{f}(\boldsymbol{r'})d\Gamma \quad (2)$$

where $k$ is the free space wave number.

When the dielectric profile of $\Omega_h$ is homogeneous, i.e. $\epsilon_h(\boldsymbol{r}) = \epsilon_h$, we denote with $k_h$ the associated wave number in $\Omega_h$. In this case, with identical definitions with respect to the ones in (1) and (2), we define the operators $\mathcal{S}_{\Gamma_h}^k$, $\mathcal{D}_{\Gamma_h}^k$ and $\mathcal{S}_{\Gamma_h}^{k_h}$, $\mathcal{D}_{\Gamma_h}^{k_h}$ for which the integrals are defined on the surface $\Gamma_h$ and for which the wave numbers are set equal to $k$ and $k_h$, respectively. In the following we will also need the definition of the surface operators $\mathcal{T}_{\Gamma_h}^k = \hat{\boldsymbol{n}} \times \mathcal{S}_{\Gamma_h}^k$, $\mathcal{T}_{\Gamma_h}^{k_h} = \hat{\boldsymbol{n}} \times \mathcal{S}_{\Gamma_h}^{k_h}$, $\mathcal{K}_{\Gamma_h}^k = \hat{\boldsymbol{n}} \times \mathcal{D}_{\Gamma_h}^{k_h}$ and $\mathcal{K}_{\Gamma_h}^{k_h} = \hat{\boldsymbol{n}} \times \mathcal{D}_{\Gamma_h}^{k_h}$ for an outward directed normal $\hat{\boldsymbol{n}}$ on the surface $\Gamma_h$ and the definition of the free space and relative wave impedance $\eta = \sqrt{\mu/\epsilon}$ and $\eta_r = k/k_h$, respectively.

When the dielectric profile of $\Omega_h$ is inhomogeneous, i.e. for general, potentially position-dependent, $\epsilon_h(\boldsymbol{r})$, we define the dielectric contrast ratio as $\chi(\boldsymbol{r}) = 1 - \frac{\epsilon_h(\boldsymbol{r})}{\epsilon}$. In this case, we define the volume operator

$$\begin{aligned}\mathcal{S}^k_{\Omega_h}(\boldsymbol{f}(\boldsymbol{r})) = ik \int_{\Omega_h} \frac{e^{ik|\boldsymbol{r}-\boldsymbol{r}'|}}{4\pi|\boldsymbol{r}-\boldsymbol{r}'|}(\chi(\boldsymbol{r}')\boldsymbol{f}(\boldsymbol{r}'))d\Omega \\ - \frac{1}{ik}\nabla \int_{\Omega_h} \frac{e^{ik|\boldsymbol{r}-\boldsymbol{r}'|}}{4\pi|\boldsymbol{r}-\boldsymbol{r}'|}\nabla' \cdot (\chi(\boldsymbol{r}')\boldsymbol{f}(\boldsymbol{r}'))d\Omega.\end{aligned} \quad (3)$$

## 3. The Inverse Source Scheme

It is very well-known from potential theory [55] that we can represent the field radiated *outside* $\Gamma$ from sources located *inside* $\Gamma$ as

$$\boldsymbol{E}_{ext}(\boldsymbol{r}) = \alpha \mathcal{S}^k_\Gamma(\boldsymbol{J}(\boldsymbol{r}', \alpha, \beta)) + \beta \mathcal{D}^k_\Gamma(\boldsymbol{M}(\boldsymbol{r}', \alpha, \beta)) \quad (4)$$

and

$$\eta \boldsymbol{H}_{ext} = \frac{1}{ik}\nabla \times \boldsymbol{E}_{ext}(\boldsymbol{r}) \quad (5)$$

where $\alpha$ and $\beta$ can be arbitrarily chosen. The values of $\boldsymbol{J}(\boldsymbol{r}')$ and $\boldsymbol{M}(\boldsymbol{r}')$ will be a function of this arbitrary choice while the resulting value of $\boldsymbol{E}_{ext}(\boldsymbol{r})$ will be independent of it. A common choice is $\alpha = \beta = 1$ for which $\boldsymbol{J}(\boldsymbol{r}')$ and $\boldsymbol{M}(\boldsymbol{r}')$ have the particularly physical meaning of being the Love's currents [56] equal to $\hat{\boldsymbol{n}}_r \times \boldsymbol{H}_{ext}(\boldsymbol{r})$ and $-\hat{\boldsymbol{n}}_r \times \boldsymbol{E}_{ext}(\boldsymbol{r})$ respectively ($\hat{\boldsymbol{n}}_r$ represents the outward directed normal on $\Gamma$). Given that we want to avoid the measurement of the magnetic field, however, this choice does not suit our treatment. We select instead $\alpha = 1$ and $\beta = 0$, in this case we deal only with the current distribution $\boldsymbol{J}(\boldsymbol{r}')$ that in the general case however, does not have a straightforward physical interpretation. We thus *define* $\boldsymbol{J}(\boldsymbol{r}')$ the (unknown) current distribution on $\Gamma$ such that

$$\boldsymbol{E}_{ext}(\boldsymbol{r}) = \mathcal{S}^k_\Gamma(\boldsymbol{J}(\boldsymbol{r}')) \quad (6)$$

and

$$\eta \boldsymbol{H}_{ext}(\boldsymbol{r}) = \frac{1}{ik}\nabla \times \mathcal{S}^k_\Gamma(\boldsymbol{J}(\boldsymbol{r}')) \quad (7)$$

$\forall \boldsymbol{r} \in \mathbb{R}^3/\Omega$. The reader should again notice that the current $\boldsymbol{J}(\boldsymbol{r}')$ will not be a simple function of the tangential component of the total magnetic field, like the standard electric current is, but it will be a more complicated function of tangential components of both total electric and magnetic fields. Given however that the radiated fields are to be recovered only in the external region of the equivalent surface, the complicated relationship between the current and the total fields will never be used nor required to be made explicit.

*3.1. Homogeneous Head Model*

We assume in this section that the dielectric profile of the head is homogeneous in $\Omega_h$. The purpose of the dosimetric assessment is to find the value of $\boldsymbol{E}(\boldsymbol{r})$, $\boldsymbol{r} \in \Omega_h$, i.e. the total electric field within the head. This field is computed using the surface currents on the surface $\Gamma_h$. These unknown surface currents on the head due to the external field radiated by the equivalent current $\boldsymbol{J}(\boldsymbol{r})$, given the homogeneity of the dielectric profile on $\Gamma_h$ can be obtained by solving the PMCHWT integral equation given by

$$\begin{bmatrix} (\mathcal{T}_{\Gamma_h}^k + \mathcal{T}_{\Gamma_h}^{k_h}/\eta_r) & -(\mathcal{K}_{\Gamma_h}^k + \mathcal{K}_{\Gamma_h}^{k_h}) \\ (\mathcal{K}_{\Gamma_h}^k + \mathcal{K}_{\Gamma_h}^{k_h}) & (\mathcal{T}_{\Gamma_h}^k + \eta_r \mathcal{T}_{\Gamma_h}^{k_h}) \end{bmatrix} \begin{bmatrix} \boldsymbol{M}_h(\boldsymbol{r}') \\ \boldsymbol{J}_h(\boldsymbol{r}') \end{bmatrix} = \begin{bmatrix} -\hat{\boldsymbol{n}} \times \eta \boldsymbol{H}_{ext}(\boldsymbol{r}) \\ -\hat{\boldsymbol{n}} \times \boldsymbol{E}_{ext}(\boldsymbol{r}) \end{bmatrix}, \quad (8)$$

where the unknown electric and magnetic surface currents denoted by $\boldsymbol{J}_h(\boldsymbol{r}')$ and $\boldsymbol{M}_h(\boldsymbol{r}')$ for $\boldsymbol{r}' \in \Gamma_h$ respectively can be found using

$$\begin{bmatrix} \boldsymbol{M}_h(\boldsymbol{r}') \\ \boldsymbol{J}_h(\boldsymbol{r}') \end{bmatrix} = \begin{bmatrix} (\mathcal{T}_{\Gamma_h}^k + \mathcal{T}_{\Gamma_h}^{k_h}/\eta_r) & -(\mathcal{K}_{\Gamma_h}^k + \mathcal{K}_{\Gamma_h}^{k_h}) \\ (\mathcal{K}_{\Gamma_h}^k + \mathcal{K}_{\Gamma_h}^{k_h}) & (\mathcal{T}_{\Gamma_h}^k + \eta_r \mathcal{T}_{\Gamma_h}^{k_h}) \end{bmatrix}^{-1} \begin{bmatrix} -\hat{\boldsymbol{n}} \times \eta \boldsymbol{H}_{ext}(\boldsymbol{r}) \\ -\hat{\boldsymbol{n}} \times \boldsymbol{E}_{ext}(\boldsymbol{r}) \end{bmatrix}, \quad (9)$$

and from which the field $\boldsymbol{E}_h(\boldsymbol{r})$ scattered by the head can be obtained as

$$\boldsymbol{E}_h(\boldsymbol{r}) = \mathcal{S}_{\Gamma_h}^k(\boldsymbol{J}_h(\boldsymbol{r}')) + \mathcal{D}_{\Gamma_h}^k(\boldsymbol{M}_h(\boldsymbol{r}')) \quad \boldsymbol{r} \in \mathbb{R}^3/\Omega_h. \quad (10)$$

or

$$\boldsymbol{E}_h(\boldsymbol{r}) = \begin{bmatrix} \mathcal{S}_{\Gamma_h}^k & \mathcal{D}_{\Gamma_h}^k \end{bmatrix} \begin{bmatrix} (\mathcal{T}_{\Gamma_h}^k + \mathcal{T}_{\Gamma_h}^{k_h}/\eta_r) & -(\mathcal{K}_{\Gamma_h}^k + \mathcal{K}_{\Gamma_h}^{k_h}) \\ (\mathcal{K}_{\Gamma_h}^k + \mathcal{K}_{\Gamma_h}^{k_h}) & (\mathcal{T}_{\Gamma_h}^k + \eta_r \mathcal{T}_{\Gamma_h}^{k_h}) \end{bmatrix}^{-1} \begin{bmatrix} -\hat{\boldsymbol{n}} \times \eta \boldsymbol{H}_{ext}(\boldsymbol{r}) \\ -\hat{\boldsymbol{n}} \times \boldsymbol{E}_{ext}(\boldsymbol{r}) \end{bmatrix}. \quad (11)$$

This can be further written as a function of $\boldsymbol{J}(\boldsymbol{r}')$ on $\Gamma$ by leveraging on (6) and (7) as

$$\boldsymbol{E}_h(\boldsymbol{r}) = \begin{bmatrix} \mathcal{S}_{\Gamma_h}^k & \mathcal{D}_{\Gamma_h}^k \end{bmatrix} \begin{bmatrix} (\mathcal{T}_{\Gamma_h}^k + \mathcal{T}_{\Gamma_h}^{k_h}/\eta_r) & -(\mathcal{K}_{\Gamma_h}^k + \mathcal{K}_{\Gamma_h}^{k_h}) \\ (\mathcal{K}_{\Gamma_h}^k + \mathcal{K}_{\Gamma_h}^{k_h}) & (\mathcal{T}_{\Gamma_h}^k + \eta_r \mathcal{T}_{\Gamma_h}^{k_h}) \end{bmatrix}^{-1} \cdot$$
$$\cdot \begin{bmatrix} -\hat{\boldsymbol{n}} \times \eta \frac{1}{ik} \nabla \times \mathcal{S}_{\Gamma}^k(\boldsymbol{J}(\boldsymbol{r}')) \\ -\hat{\boldsymbol{n}} \times \mathcal{S}_{\Gamma}^k(\boldsymbol{J}(\boldsymbol{r}')) \end{bmatrix}. \quad (12)$$

By enforcing the condition $\boldsymbol{E}(\boldsymbol{r}) = \boldsymbol{E}_{ext}(\boldsymbol{r}) + \boldsymbol{E}_h(\boldsymbol{r})$ on $\Gamma_m$ we obtain

$$\mathcal{S}_{\Gamma}^k(\boldsymbol{J}(\boldsymbol{r}')) + \begin{bmatrix} \mathcal{S}_{\Gamma_h}^k & \mathcal{D}_{\Gamma_h}^k \end{bmatrix} \begin{bmatrix} (\mathcal{T}_{\Gamma_h}^k + \mathcal{T}_{\Gamma_h}^{k_h}/\eta_r) & -(\mathcal{K}_{\Gamma_h}^k + \mathcal{K}_{\Gamma_h}^{k_h}) \\ (\mathcal{K}_{\Gamma_h}^k + \mathcal{K}_{\Gamma_h}^{k_h}) & (\mathcal{T}_{\Gamma_h}^k + \eta_r \mathcal{T}_{\Gamma_h}^{k_h}) \end{bmatrix}^{-1} \cdot$$
$$\cdot \begin{bmatrix} -\hat{\boldsymbol{n}} \times \eta \frac{1}{ik} \nabla \times \mathcal{S}_{\Gamma}^k(\boldsymbol{J}(\boldsymbol{r}')) \\ -\hat{\boldsymbol{n}} \times \mathcal{S}_{\Gamma}^k(\boldsymbol{J}(\boldsymbol{r}')) \end{bmatrix} = \boldsymbol{E}(\boldsymbol{r}), \quad \boldsymbol{r} \in \Gamma_m. \quad (13)$$

As pointed out in the previous section, the field $\boldsymbol{E}(\boldsymbol{r})$ is assumed to be known as a result of a measurement procedure, so that the unknown current distribution $\boldsymbol{J}(\boldsymbol{r}')$ can be obtained by inverting (13). This inversion will be obtained by subsequent use of boundary element discretizations as explained in the following section.

*3.1.1. Discretization*

In order to solve numerically the integral equations defined in the previous section, it is necessary to discretize the involved surface operators. As in any standard boundary element method, we need to approximate the various geometries present with a triangular tessellation giving rise to surface meshes for $\Gamma_h$, $\Gamma$ and $\Gamma_m$. On the internal edges of these surface meshes, we can define the sets of Rao-Wilton-Glisson (RWG) basis functions [57] $\{\boldsymbol{f}_n^h\}_{n=1}^{N_h}$, $\{\boldsymbol{f}_n\}_{n=1}^{N}$ and $\{\boldsymbol{f}_n^m\}_{n=1}^{N_m}$ respectively ($N_h$, $N$ and $N_m$ being the number of edges on $\Gamma_h$, $\Gamma$ and $\Gamma_m$ respectively). The surface currents can be approximated as

$$\boldsymbol{J}_h(\boldsymbol{r}) = \sum_{n=1}^{N_h} \alpha_n \boldsymbol{f}_n^h(\boldsymbol{r})$$
$$\boldsymbol{M}_h(\boldsymbol{r}) = \sum_{n=1}^{N_h} \beta_n \boldsymbol{f}_n^h(\boldsymbol{r}) \qquad (14)$$
$$\boldsymbol{J}(\boldsymbol{r}) = \sum_{n=1}^{N} \gamma_n \boldsymbol{f}_n(\boldsymbol{r}).$$

The different blocks in equation (13) can be discretized as explained in what follows. The PMCHWT operators $\mathcal{T}_{\Gamma_h}^k$, $\mathcal{T}_{\Gamma_h}^{k_h}$, $\mathcal{K}_{\Gamma_h}^k$ and $\mathcal{K}_{\Gamma_h}^{k_h}$ are discretized using the sets of RWG basis functions both as source and testing functions resulting in the matrices

$$\begin{aligned}
(\mathbf{T})_{mn} &= \left\langle \hat{\boldsymbol{n}} \times \boldsymbol{f}_m^h(\boldsymbol{r}), \mathcal{T}_{\Gamma_h}^k(\boldsymbol{f}_n^h(\boldsymbol{r})) \right\rangle_{\Gamma_h}, \\
(\mathbf{T}')_{mn} &= \left\langle \hat{\boldsymbol{n}} \times \boldsymbol{f}_m^h(\boldsymbol{r}), \mathcal{T}_{\Gamma_h}^{k_h}(\boldsymbol{f}_n^h(\boldsymbol{r})) \right\rangle_{\Gamma_h}, \\
(\mathbf{K})_{mn} &= \left\langle \hat{\boldsymbol{n}} \times \boldsymbol{f}_m^h(\boldsymbol{r}), \mathcal{K}_{\Gamma_h}^k(\boldsymbol{f}_n^h(\boldsymbol{r})) \right\rangle_{\Gamma_h}, \\
(\mathbf{K}')_{mn} &= \left\langle \hat{\boldsymbol{n}} \times \boldsymbol{f}_m^h(\boldsymbol{r}), \mathcal{K}_{\Gamma_h}^{k_h}(\boldsymbol{f}_n^h(\boldsymbol{r})) \right\rangle_{\Gamma_h}
\end{aligned} \qquad (15)$$

where the notation $\langle \boldsymbol{a} \cdot \boldsymbol{b} \rangle_\chi = \int_\chi \boldsymbol{a} \cdot \boldsymbol{b}\, d\chi$ and $(\mathbf{A})_{mn}$ represents the element in the row $m$ and the column $n$ of a matrix $\mathbf{A}$. The discretized version of the continuous operators applied to the equivalent source currents radiating the fields on the head surface are given by

$$\begin{aligned}
(\mathbf{K}_\Gamma)_{mn} &= \left\langle \hat{\boldsymbol{n}} \times \boldsymbol{f}_m^h(\boldsymbol{r}), -\hat{\boldsymbol{n}} \times \eta_0 \frac{1}{ik} \nabla \times \mathcal{S}_\Gamma^k(\boldsymbol{f}_n(\boldsymbol{r})) \right\rangle_{\Gamma_h}, \\
(\mathbf{T}_\Gamma)_{mn} &= \left\langle \hat{\boldsymbol{n}} \times \boldsymbol{f}_m^h(\boldsymbol{r}), -\hat{\boldsymbol{n}} \times \mathcal{S}_\Gamma^k(\boldsymbol{f}_n(\boldsymbol{r})) \right\rangle_{\Gamma_h}.
\end{aligned} \qquad (16)$$

Similarly, the discretized version of the continuous operators applied to the surface currents on the head radiating the fields on the measurement surface

are given by

$$(\mathbf{S}_{\Gamma_h})_{mn} = \left\langle \boldsymbol{f}_m^m(\boldsymbol{r}), \mathcal{S}_{\Gamma_h}^k(\boldsymbol{f}_n^h(\boldsymbol{r})) \right\rangle_\Gamma,$$
$$(\mathbf{D}_{\Gamma_h})_{mn} = \left\langle \boldsymbol{f}_m^m(\boldsymbol{r}), \mathcal{D}_{\Gamma_h}^k(\boldsymbol{f}_n^h(\boldsymbol{r})) \right\rangle_\Gamma. \tag{17}$$

The incident electric field due to the equivalent source currents on the measurement surface represented as a matrix is given by

$$(\mathbf{S}_\Gamma)_{mn} = \left\langle \boldsymbol{f}_m^m(\boldsymbol{r}), \mathcal{S}_\Gamma^k(\boldsymbol{f}_n(\boldsymbol{r})) \right\rangle_{\Gamma_m}. \tag{18}$$

The overall right hand side vector reads

$$\mathbf{v}_m = \langle \boldsymbol{f}_m^m(\boldsymbol{r}), \boldsymbol{E}(\boldsymbol{r}) \rangle_{\Gamma_m}. \tag{19}$$

Finally, define the unknown current coefficients vector

$$(\boldsymbol{\gamma})_m = \gamma_m. \tag{20}$$

Leveraging on all the definitions above, the discretization of equation (13) reads

$$\mathbf{S}_\Gamma \boldsymbol{\gamma} + \begin{bmatrix} \mathbf{S}_{\Gamma_h} & \mathbf{D}_{\Gamma_h} \end{bmatrix} \begin{bmatrix} (\mathbf{T} + \mathbf{T}'/\eta_r) & -(\mathbf{K} + \mathbf{K}') \\ (\mathbf{K} + \mathbf{K}') & (\mathbf{T} + \eta_r \mathbf{T}') \end{bmatrix}^{-1} \begin{bmatrix} \mathbf{K}_\Gamma \\ \mathbf{T}_\Gamma \end{bmatrix} \boldsymbol{\gamma} = \mathbf{v}. \tag{21}$$

*3.2. Inhomogeneous Head Model*

In realistic scenarios, the modeling of the human head phantom may result in inhomogeneous dielectric profiles modeling different tissues (typically scalp, skull and brain). The objective of dosimetric assessments remains unchanged from the case analyzed in the previous section, i.e. the total electric field inside the head region must be recovered starting from measurements in the vicinities of the radiating source. The total electric field now is the sum of the external field radiated by the equivalent current $\boldsymbol{J}(\boldsymbol{r}')$ and the equivalent volume current $\boldsymbol{J}_h^v(\boldsymbol{r}')$ inside the human head model. The mapping between the volume currents and the external field can be expressed using a Volume Integral Equation (VIE) [58, 59]

$$\left[(1 - \chi(\boldsymbol{r}))\frac{\mathcal{I}}{ik} + \mathcal{S}_{\Omega_h}^k\right] \boldsymbol{J}_h^v(\boldsymbol{r}') = \boldsymbol{E}_{ext}(\boldsymbol{r}) \quad \boldsymbol{r} \in \Omega_h. \tag{22}$$

where $\mathcal{I}$ is the identity operator and the unknown volume current is given by

$$\boldsymbol{J}_h^v(\boldsymbol{r}) = -\left[(1 - \chi(\boldsymbol{r}))\frac{\mathcal{I}}{ik} + \mathcal{S}_{\Omega_h}^k\right]^{-1} \boldsymbol{E}_{ext}(\boldsymbol{r}') \quad \boldsymbol{r} \in \Omega_h. \tag{23}$$

The electric field in the external region due to the volume current inside the head is given by

$$\boldsymbol{E}_h(\boldsymbol{r}) = -\mathcal{S}_{\Omega_h}^k \boldsymbol{J}_h^v(\boldsymbol{r}') \quad \boldsymbol{r} \in \mathbb{R}^3/\Omega_h. \tag{24}$$

or

$$\boldsymbol{E}_h(\boldsymbol{r}) = -\mathcal{S}_{\Omega_h}^k \left[ (1-\chi(\boldsymbol{r}))\frac{\mathcal{I}}{ik} + \mathcal{S}_{\Omega_h}^k \right]^{-1} \boldsymbol{E}_{ext}(\boldsymbol{r}'). \tag{25}$$

which can be written in terms of surface currents on the equivalent surface as

$$\boldsymbol{E}_h(\boldsymbol{r}) = -\mathcal{S}_{\Omega_h}^k \left[ (1-\chi(\boldsymbol{r}))\frac{\mathcal{I}}{ik} + \mathcal{S}_{\Omega_h}^k \right]^{-1} \mathcal{S}_\Gamma^k(\boldsymbol{J}(\boldsymbol{r}')). \tag{26}$$

The total electric field on the measurement surface $\Gamma_m$ is thus obtained as

$$\mathcal{S}_\Gamma^k(\boldsymbol{J}(\boldsymbol{r}')) - \mathcal{S}_{\Omega_h}^k \left[ (1-\chi(\boldsymbol{r}))\frac{\mathcal{I}}{ik} + \mathcal{S}_{\Omega_h}^k \right]^{-1} \mathcal{S}_\Gamma^k(\boldsymbol{J}(\boldsymbol{r}')) = \boldsymbol{E}(\boldsymbol{r}) \quad \boldsymbol{r} \in \Gamma_m. \tag{27}$$

The above equation gives the relationship between the unknown equivalent surface current and the measured electric field in the presence of an inhomogeneous head model. The unknown surface current $\boldsymbol{J}(\boldsymbol{r}')$ can be obtained by solving the discretized version of the above equation as explained in the next subsection.

*3.2.1. Discretization*

Equation (27) can be discretized after meshing the equivalent surface $\Gamma$ with triangular cells and the head volume $\Omega_h$ with tetrahedral cells. On a pair of triangular cells of the equivalent source and measurement surface, we define the sets of RWG basis functions $\{\boldsymbol{f}_n\}_{n=1}^N$ and $\{\boldsymbol{f}_n^m\}_{n=1}^{N_m}$ respectively. Similarly, on a pair of each adjacent tetrahedron of the volume mesh, we define an Schaubert-Wilton-Glisson (SWG) basis function [58] giving rise to the set $\{\boldsymbol{f}_n^v\}_{n=1}^{N_v}$ ($N_v$ represents the number of triangular cells based on which the SWG basis functions are defined). Using these basis functions, we can discretize the volume and surface currents as

$$\begin{aligned} \boldsymbol{J}_v(\boldsymbol{r}) &= \sum_{n=1}^{N_v} \alpha_n \boldsymbol{f}_n^v(\boldsymbol{r}) \\ \boldsymbol{J}(\boldsymbol{r}) &= \sum_{n=1}^{N} \gamma_n \boldsymbol{f}_n(\boldsymbol{r}). \end{aligned} \tag{28}$$

The VIE operator can be discretized as

$$(\mathbf{S}_{\Omega_h})_{mn} = \langle \boldsymbol{f}_m^v(\boldsymbol{r}), \mathcal{S}_{\Omega_h}^k(\boldsymbol{f}_n^v(\boldsymbol{r})) \rangle_{\Omega_h} \tag{29}$$

and the Gram matrix is given by

$$(\mathbf{G}_{\Omega_h})_{mn} = \left\langle \boldsymbol{f}_m^v(\boldsymbol{r}), \frac{1-\chi(\boldsymbol{r})}{ik} \boldsymbol{f}_n^v(\boldsymbol{r}) \right\rangle_{\Omega_h}. \tag{30}$$

The discretized surface operator applied to the equivalent source current radiating on the head surface is given by

$$(\mathbf{S}_\Gamma^{\Omega_h})_{mn} = \langle \boldsymbol{f}_m^v(\boldsymbol{r}), \mathcal{S}_\Gamma^k(\boldsymbol{f}_n(\boldsymbol{r})) \rangle_{\Omega_h}. \tag{31}$$

Similarly, the discretized volume operator applied to the volume currents of the inhomogeneous head model radiating on the measurement surface is given by

$$(\mathbf{S}^{\Gamma}_{\Omega_h})_{mn} = \left\langle \boldsymbol{f}_m(\boldsymbol{r}), \mathcal{S}^k_{\Omega_h}(\boldsymbol{f}^v_n(\boldsymbol{r})) \right\rangle_{\Gamma_m}. \tag{32}$$

The discretization of the operator defined for the incident electric field due to the equivalent source currents on the measurement surface

$$(\mathbf{S}_{\Gamma})_{mn} = \left\langle \boldsymbol{f}_m(\boldsymbol{r}), \mathcal{S}^k_{\Gamma}(\boldsymbol{f}_n(\boldsymbol{r})) \right\rangle_{\Gamma_m}. \tag{33}$$

and the right hand side and solution vectors

$$\mathbf{v}_m = \left\langle \boldsymbol{f}^m_m(\boldsymbol{r}), \boldsymbol{E}(\boldsymbol{r}) \right\rangle_{\Gamma_m}. \tag{34}$$

$$(\boldsymbol{\gamma})_m = \gamma_m. \tag{35}$$

remain similar to those of the homogeneous case (treated in the previous section). By combining the discretizations above we obtain the following discretized version of equation (27)

$$\mathbf{S}_{\Gamma}\gamma - \mathbf{S}^{\Gamma}_{\Omega_h} \left[ \mathbf{G}_{\Omega_h} + \mathbf{S}_{\Omega_h} \right]^{-1} \mathbf{S}^{\Omega_h}_{\Gamma} \gamma = \mathbf{v}. \tag{36}$$

## 4. Calderón preconditioning

As delineated in the previous sections, the linear system of equations (21) and (36) requires near field measurements for building up the right hand side vector. A practical case of interest arises when the equivalent surface and the measurement surface coincide i.e. $\Gamma_m = \Gamma$. Even in this case, however, the surface meshes on $\Gamma_m$ and $\Gamma$ are often different. In fact, the discretization of the measurement surface is dependent on the overall number of degrees of freedom of the measured near field values [19, 20] and also on the limitations of the measurement system setups [21, 60]. On the other hand, the discretization of the equivalent surface depends on how well it can model the near field behaviour of the radiating source and it plays a role in the pseudo-inversion procedure. This results in a different discretization density on the equivalent surface compared to the measurement surface. Therefore, the system matrices in equation (21) and (36) are rectangular and they deal with unequal number of unknowns compared to the number of equations. These equations can be solved in the least squares sense using an iterative solver by applying the transpose to the matrix $\mathbf{S}_{\Gamma}$ as

$$\mathbf{S}^T_{\Gamma}[\mathbf{S}_{\Gamma} + \tilde{\mathbf{S}}_{\Gamma}]\gamma = \mathbf{S}^T_{\Gamma}\mathbf{v}. \tag{37}$$

where in case of the homogeneous head model

$$\tilde{\mathbf{S}}_{\Gamma} = \begin{bmatrix} \mathbf{S}_{\Gamma_h} & \mathbf{D}_{\Gamma_h} \end{bmatrix} \begin{bmatrix} (\mathbf{T} + \mathbf{T}'/\eta_r) & -(\mathbf{K} + \mathbf{K}') \\ (\mathbf{K} + \mathbf{K}') & (\mathbf{T} + \eta_r \mathbf{T}') \end{bmatrix}^{-1} \begin{bmatrix} \mathbf{S}_{\Gamma} \\ \mathbf{K}_{\Gamma} \end{bmatrix}, \tag{38}$$

and for the inhomogeneous head model

$$\tilde{\mathbf{S}}_\Gamma = -\mathbf{S}_{\Omega_h}^\Gamma [\mathbf{G}_{\Omega_h} + \mathbf{S}_{\Omega_h}]^{-1} \mathbf{S}_\Gamma^{\Omega_h}. \tag{39}$$

In equation (37), the matrix $\tilde{\mathbf{S}}_\Gamma$ represents a compact perturbation of the EFIE operator matrix with a negligible contribution to the spectrum of $\mathbf{S}_\Gamma$. The overall system matrix inherits the ill-conditioning of the EFIE operator due to the dense discretization and low frequency breakdown [61]. The condition number of $\mathbf{S}_\Gamma$ shows a behavior of order $\mathcal{O}(\frac{1}{(kh)^2})$ [62, 63]. This results in a conditioning behaviour of order $\mathcal{O}(\frac{1}{(kh)^4})$ for the system matrix $\mathbf{S}_\Gamma^T \mathbf{S}_\Gamma$. Therefore, solving equation (37) becomes a challenging task as the discretization on the equivalent surface increases and/or the frequency decreases. In this work we propose to leverage on Calderón preconditioning to solve this problem. Calderón preconditioning is based on the Calderón identity [64, 65]

$$(\mathcal{T}_\Gamma^k)^2 (\boldsymbol{J}(\boldsymbol{r})) = -\frac{\boldsymbol{J}(\boldsymbol{r})}{4} + (\tilde{\mathcal{K}}_\Gamma^k)^2 (\boldsymbol{J}(\boldsymbol{r})). \tag{40}$$

The discretization of equation (40) results in well conditioned system matrices [64]. The operator $\mathcal{T}_\Gamma^k$ when discretized with the sets of RWG basis functions $\{\boldsymbol{f}_n\}_{n=1}^N$ and $\{\hat{\boldsymbol{n}} \times \boldsymbol{f}_n^m\}_{n=1}^{N_m}$ results in the matrix $\mathbf{S}_\Gamma$. To realize the discretization of the Calderón identity (equation (40)), however, we also need the sets of Buffa-Christiansen (BC) basis functions $\{\boldsymbol{b}_n\}_{n=1}^N$ and $\{\hat{\boldsymbol{n}} \times \boldsymbol{b}_n^m\}_{n=1}^{N_m}$ (the definition of these functions is omitted here for the sake of brevity, the reader should refer to [66] or [64] for the implementational details of these boundary elements). This discretization gives rise to the matrix $\mathbf{S}_\Gamma^{BC} = \langle \boldsymbol{b}_m^m(\boldsymbol{r}), \mathcal{S}_\Gamma(\boldsymbol{b}_n(\boldsymbol{r})) \rangle_{\Gamma_m}$. In order to link correctly the basis functions between the two operator matrices $\mathbf{S}_\Gamma^{BC}$ and $\mathbf{S}_\Gamma$, we need a suitable Gram matrix $\mathbf{G}_{\Gamma_m} = \langle \hat{\boldsymbol{n}} \times \boldsymbol{f}_m^m(\boldsymbol{r}), \boldsymbol{b}_m^m(\boldsymbol{r}) \rangle_{\Gamma_m}$. The proposed regularized linear system of equations for dosimetry assessment then reads

$$[\mathbf{S}_\Gamma^{BC}]^T [\mathbf{G}_{\Gamma_m}]^{-1} [\mathbf{S}_\Gamma + \tilde{\mathbf{S}}_\Gamma] \boldsymbol{\gamma} = [\mathbf{S}_\Gamma^{BC}]^T [\mathbf{G}_{\Gamma_m}]^{-1} \mathbf{v}. \tag{41}$$

## 5. Numerical Results

The first test focused on the validation of the BEM formulation, developed in section 3.1 for homogeneous phantom profiles. The electric field radiated by a mobile phone antenna in the presence of an homogeneous head phantom is shown in Figure 2a. The mobile antenna has been enclosed in a parallelepiped's equivalent surface that is also used as measurement surface. After applying the numerical procedure detailed in section 3.1, the electric field is reconstructed in Figure 2b. Figure 2c shows the reconstruction relative error computed as

$$\varepsilon(\boldsymbol{r}) = \frac{|\boldsymbol{E}(\boldsymbol{r}) - \boldsymbol{E}_r(\boldsymbol{r})|}{|\boldsymbol{E}(\boldsymbol{r})|} \tag{42}$$

where the electric field $\boldsymbol{E}(\boldsymbol{r})$ and $\boldsymbol{E}_r(\boldsymbol{r})$ are due to the radiating source and the Huygens surface in presence of the head respectively. The maximum field relative error stays below 1% confirming the validity of the formulation.

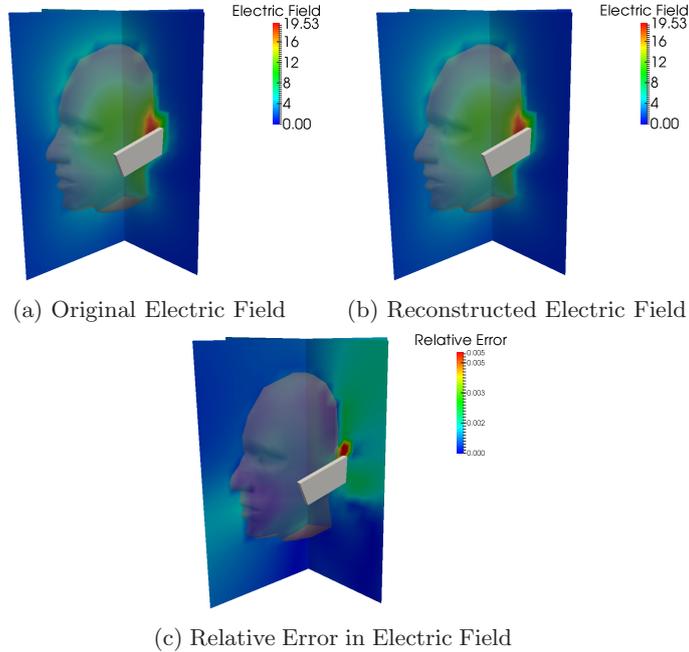

(a) Original Electric Field  (b) Reconstructed Electric Field

(c) Relative Error in Electric Field

Figure 2: Homogeneous head model

A second numerical test has been performed to assess the performance of the BEM formulation, developed in section 3.2 for inhomogeneous phantom profiles. The inhomogeneous head phantom we have used is shown in Figure 3. We have adopted an MRI based three layers head model which has a piecewise-constant dielectric profile. The reader should notice, however, that any other dielectric profile of arbitrary inhomogeneity could have been used as well. The electric field radiated by a mobile phone antenna in the presence of this inhomogeneous head phantom is shown in Figure 4a. The mobile antenna has been again enclosed in a parallelepiped's equivalent surface that is also used as measurement surface. After applying the numerical procedure detailed in section 3.2, the electric field is reconstructed in Figure 4b. The reconstruction relative error is shown in Figure 2c, the maximum field relative error stays well below 1% confirming that also this formulation for inhomogeneous dielectric profiles is a valid one.

A final set of tests has been focused in validating the Calderón regularizations proposed in section 4. First we have considered two canonical cases (a cube and an hemisphere) to check the stability of the regularization to mesh refinement. To this purpose the measurement mesh has been kept constant, while the equivalent surface mesh density was increased. For the case of the cube, the condition numbers of the overall system matrix w.r.t to the average edge length is shown in Figure 5. It is clear that the regularized system matrix has a constant condition number when the mesh density increases whereas without regularization a steep condition number growth is observed. A similar behavior is observed in the case of the hemisphere as it is shown in Figure 6. The stability of the regularized system with respect to frequency has been tested as well. The condition numbers as a function of the frequency are shown in Figure 7. Also in this case the regularized formulation is stable, while standard operators show the expected frequency instability. A regularization in a real case scenario has been equally tested. To this purpose we have used the previous homogeneous head model. The relative tolerance of a CGS solver w.r.t. number of iterations to compute the dosimetry assessment can be seen in Figure 8. The tolerance curves show that the regularization proposed here can greatly improve the convergence behaviour and, as a consequence, the time necessary for the dosimetric assessment.

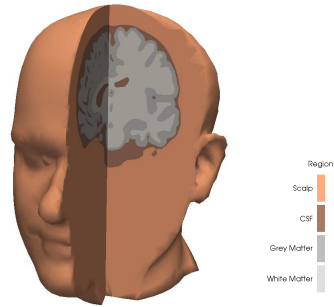

Figure 3: Volume regions of the human head model

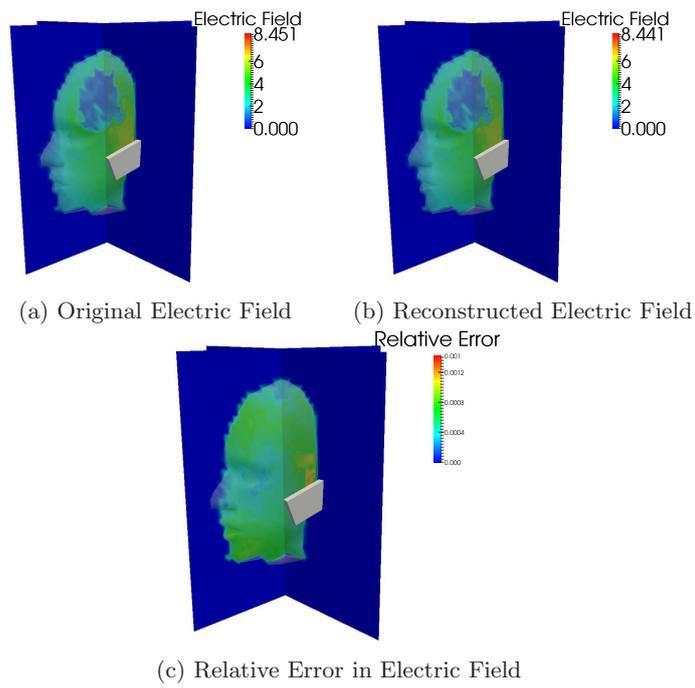

(a) Original Electric Field  (b) Reconstructed Electric Field

(c) Relative Error in Electric Field

Figure 4: Inhomogeneous head model

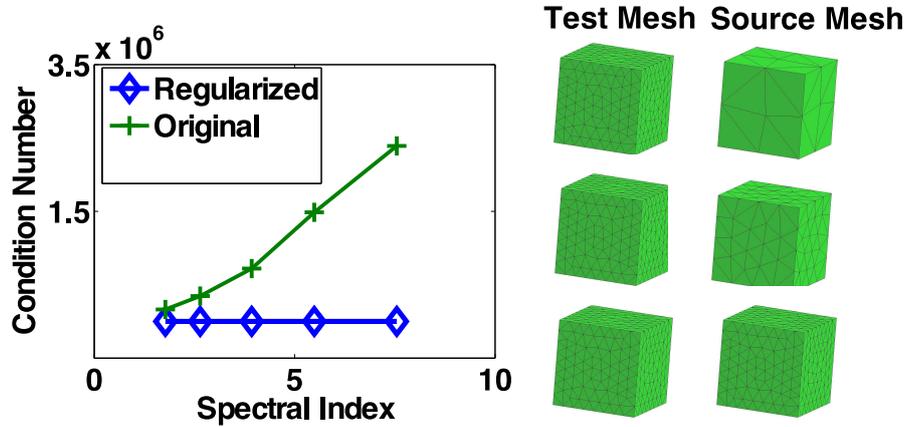

Figure 5: Variation of Condition Number of a Cube Mesh

6. Conclusions

A Boundary Element Method (BEM) formulation for contactless electromagnetic field assessments has been presented. The new scheme, which is based on a regularized BEM approach, requires the use of electric field measurements only. The regularization, which is based on Calderon techniques, enables the use of highly discretized Huygens surfaces that can be consequently placed very near to the radiating source. An hybridization with both surfacic homogeneous and volumetric inhomogeneous forward BEM solvers has been proposed that allows the use of inhomogeneous and realistic head phantoms. Numerical results have been presented that corroborate the theory and confirms the practical effectiveness of all newly proposed formulations. Future investigations will include the extension of the new formulations presented here to the time domain as well as their application to an industrial-level measurement setting.

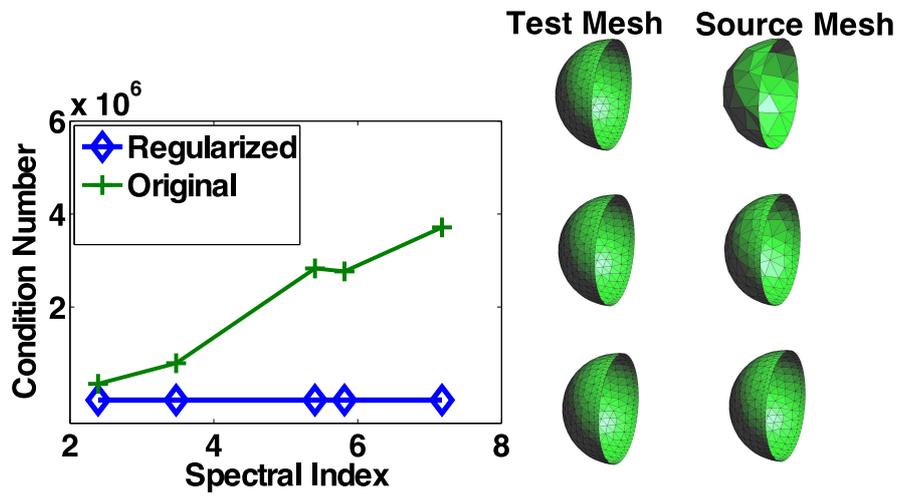

Figure 6: Variation of Condition Number of a Hemisphere Mesh

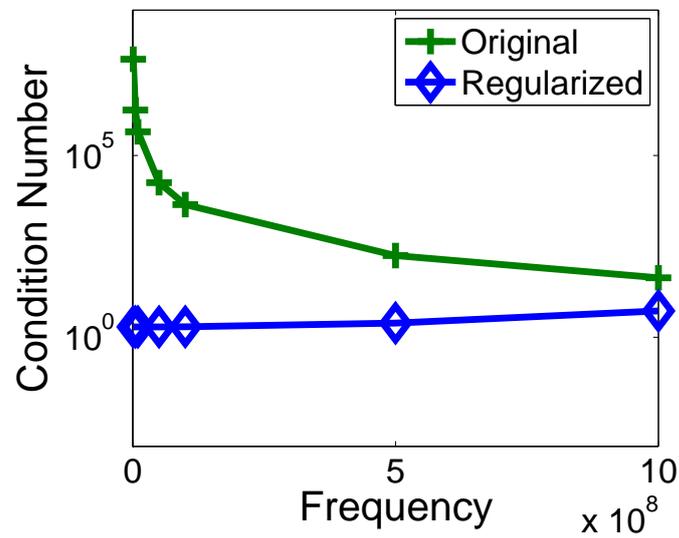

Figure 7: Variation of Condition Number of a Cube Mesh w.r.t Frequency

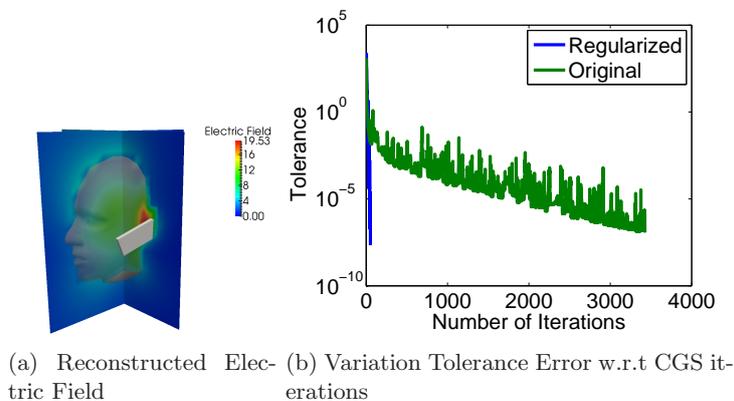

(a) Reconstructed Electric Field   (b) Variation Tolerance Error w.r.t CGS iterations

Figure 8: Validation of Calderón regularization in a real case scenario.

**Acknowledgements**

This work was supported in part by the Agence Nationale de la Recherche under Project FASTEEG-ANR-12-JS09-0010, in part by the European Union Marie Curie Project NEUROIMAGEEG, in part by the Brittany Region, France, and in part by the HPC resources through the GENCI-TGCC Project under Grant 2015-gen6944.

**References**

[1] Recommendation Council, 519/EC of 12 July 1999 on the limitation of exposure of the general public to electromagnetic fields (0 Hz to 300 GHz), Official Journal of the European Communities L 59 (1999) 59–70.

[2] M. Grandolfo, Worldwide standards on exposure to electromagnetic fields: an overview, The Environmentalist 29 (2) (2009) 109–117.

[3] J. Lin, R. Saunders, K. Schulmeister, P. Söderberg, A. Swerdlow, M. Taki, B. Veyret, G. Ziegelberger, M. H. Repacholi, R. Matthes, et al., ICNIRP Guidelines for limiting exposure to time-varying electric and magnetic fields (1 Hz to 100 kHz)., Health Physics 99 (2010) 818–836.

[4] International Commission on Non-Ionizing Radiation Protection and others, Guidance on determining compliance of exposure to pulsed and complex non-sinusoidal waveforms below 100 kHz with ICNIRP guidelines, Health physics 84 (3) (2003) 383–387.

[5] International Commission on Non-Ionizing Radiation Protection and others, ICNIRP statement on the "guidelines for limiting exposure to time-varying electric, magnetic, and electromagnetic fields (up to 300 GHz)", Health Physics 97 (3) (2009) 257–258.

[6] C. Chou, H. Bässen, J. Osepchuk, G. Balzano, R. Petersen, M. Meitz, R. Cleveland, J. Lin, L. Heynick, Radio frequency electromagnetic exposure, Journal of Bioelectromagnetics 17 (1996) 195–208.

[7] T. Schmid, O. Egger, N. Kuster, Automated E-field scanning system for dosimetric assessments, IEEE Transactions on Microwave Theory and Techniques 44 (1) (1996) 105–113.

[8] Q. Balzano, O. Garay, T. J. Manning Jr, Electromagnetic energy exposure of simulated users of portable cellular telephones, IEEE Transactions on Vehicular Technology 44 (3) (1995) 390–403.

[9] International Electrotechnical Commission and others, Human exposure to radio frequency fields from hand-held and body-mounted wireless communication devices-human models, instrumentation, and procedures, 2009.

[10] IEEE Recommended Practice for Determining the Peak Spatial-Average Specific Absorption Rate (SAR) in the Human Head from Wireless Communications Devices: Measurement Techniques, IEEE Std 1528-2003 (2003) 1–120 `doi:10.1109/IEEESTD.2003.94414`.

[11] CENELEC EN50383, Basic standard for the calculation and measurement of electromagnetic field strength and SAR related to human exposure from radio base stations and fixed terminal stations for wireless telecommunication systems (110 MHz–40 GHz), Technical committee 211.

[12] IEEE Recommended Practice for Determining the Peak Spatial-Average Specific Absorption Rate (SAR) in the Human Head from Wireless Communications Devices: Measurement Techniques - Amendment 1: CAD File for Human Head Model (SAM Phantom), IEEE Std 1528a-2005 (Amendment to IEEE Std 1528-2003) (2006) 1–11 `doi:10.1109/IEEESTD.2006.99106`.

[13] P. Glover, R. Bowtell, Measurement of electric fields due to time-varying magnetic field gradients using dipole probes, Physics in medicine and biology 52 (17) (2007) 5119.

[14] C. Gottueb, M. Hagmann, T. Babü, A. Abitbol, A. Lewin, P. Houdek, J. Schwade, Interstitial microwave hyperthermia applicators having submillimetre diameters, International journal of hyperthermia 6 (3) (1990) 707–714.

[15] C. Gabriel, S. Gabriel, E. Corthout, The dielectric properties of biological tissues: I. Literature survey, Physics in medicine and biology 41 (11) (1996) 2231.

[16] S. Gabriel, R. Lau, C. Gabriel, The dielectric properties of biological tissues: II. Measurements in the frequency range 10 Hz to 20 GHz, Physics in medicine and biology 41 (11) (1996) 2251.

[17] B. B. Beard, W. Kainz, T. Onishi, T. Iyama, S. Watanabe, O. Fujiwara, J. Wang, G. Bit-Babik, A. Faraone, J. Wiart, et al., Comparisons of computed mobile phone induced SAR in the SAM phantom to that in anatomically correct models of the human head, Electromagnetic Compatibility, IEEE Transactions on 48 (2) (2006) 397–407.

[18] J. Wiart, A. Hadjem, M. Wong, I. Bloch, Analysis of RF exposure in the head tissues of children and adults, Physics in medicine and biology 53 (13) (2008) 3681.

[19] O. M. Bucci, C. Gennarelli, C. Savarese, Representation of electromagnetic fields over arbitrary surfaces by a finite and nonredundant number of samples, IEEE Transactions on Antennas and Propagation 46 (3) (1998) 351–359.

[20] O. M. Bucci, G. Franceschetti, On the spatial bandwidth of scattered fields, IEEE Transactions on Antennas and Propagation 35 (12) (1987) 1445–1455.

[21] O. Bucci, T. Isernia, Electromagnetic inverse scattering: Retrievable information and measurement strategies, Radio Science 32 (6) (1997) 2123–2137.

[22] O. M. Bucci, L. Crocco, T. Isernia, V. Pascazio, Subsurface inverse scattering problems: quantifying, qualifying, and achieving the available information, IEEE Transactions on Geoscience and Remote Sensing 39 (11) (2001) 2527–2538.

[23] A. Hadjem, D. Lautru, C. Dale, M. F. Wong, V. F. Hanna, J. Wiart, Study of specific absorption rate (SAR) induced in two child head models and in adult heads using mobile phones, Microwave Theory and Techniques, IEEE Transactions on 53 (1) (2005) 4–11.

[24] M.-F. Wong, J. Wiart, Modelling of electromagnetic wave interactions with the human body, Comptes Rendus Physique 6 (6) (2005) 585–594.

[25] E. Conil, A. Hadjem, F. Lacroux, M. Wong, J. Wiart, Variability analysis of SAR from 20 MHz to 2.4 GHz for different adult and child models using finite-difference time-domain, Physics in medicine and biology 53 (6) (2008) 1511.

[26] G. Scarella, O. Clatz, S. Lanteri, G. Beaume, S. Oudot, J.-P. Pons, S. Piperno, P. Joly, J. Wiart, Realistic numerical modelling of human head tissue exposure to electromagnetic waves from cellular phones, Comptes Rendus Physique 7 (5) (2006) 501–508.

[27] M. Mashevich, D. Folkman, A. Kesar, A. Barbul, R. Korenstein, E. Jerby, L. Avivi, Exposure of human peripheral blood lymphocytes to electromagnetic fields associated with cellular phones leads to chromosomal instability, Bioelectromagnetics 24 (2) (2003) 82–90.

[28] J. Bakker, M. Paulides, E. Neufeld, A. Christ, X. Chen, N. Kuster, G. Van Rhoon, Children and adults exposed to low-frequency magnetic fields at the icnirp reference levels: theoretical assessment of the induced electric fields, Physics in medicine and biology 57 (7) (2012) 1815.

[29] G. Lazzi, O. P. Gandhi, On modeling and personal dosimetry of cellular telephone helical antennas with the FDTD code, IEEE Transactions on Antennas and Propagation 46 (4) (1998) 525–530.

[30] Y. Liu, Z. Liang, Z. Yang, Computation of electromagnetic dosimetry for human body using parallel FDTD algorithm combined with interpolation technique, Progress In Electromagnetics Research 82 (2008) 95–107.

[31] F. J. Meyer, D. B. Davidson, U. Jakobus, M. A. Stuchly, Human exposure assessment in the near field of GSM base-station antennas using a hybrid finite element/method of moments technique, IEEE Transactions on Biomedical Engineering 50 (2) (2003) 224–233.

[32] K. D. Paulsen, X. Jia, J. M. Sullivan Jr, Finite element computations of specific absorption rates in anatomically conforming full-body models for hyperthermia treatment analysis, IEEE Transactions on Biomedical Engineering 40 (9) (1993) 933–945.

[33] J. Hand, Modelling the interaction of electromagnetic fields (10 MHz - 10 GHz) with the human body: methods and applications, Physics in medicine and biology 53 (16) (2008) R243.

[34] D. Lautru, J. Wiart, W. Tabbara, Calculation of the power deposited in a phantom close to a base station antenna using a hybrid fdtd-momtd approach, in: Microwave Conference, 2000. 30th European, IEEE, 2000, pp. 1–4.

[35] Y. Pinto, A. Ghanmi, A. Hadjem, E. Conil, T. Namur, C. Person, J. Wiart, Numerical mobile phone models validated by sar measurements, in: Proceedings of the 5th European Conference on Antennas and Propagation (EUCAP), 2011.

[36] J. Zhu, D. Jiao, A theoretically rigorous full-wave finite-element-based solution of Maxwell's equations from dc to high frequencies, IEEE Transactions on Advanced Packaging 33 (4) (2010) 1043–1050.

[37] O. Bottauscio, M. Chiampi, L. Zilberti, A boundary element approach to relate surface fields with the specific absorption rate (SAR) induced in 3-D human phantoms, Engineering Analysis with Boundary Elements 35 (4) (2011) 657–666.

[38] O. Bottauscio, M. Chiampi, L. Zilberti, Boundary element approaches for the evaluation of human exposure to low frequency electromagnetic fields, IEEE transactions on magnetics 45 (3) (2009) 1674–1677.

[39] D. Giordano, L. Zilberti, M. Borsero, M. Chiampi, O. Bottauscio, Experimental validation of MRI dosimetric simulations in phantoms including metallic objects, IEEE Transactions on Magnetics 50 (11) (2014) 1–4.

[40] R. Albanese, P. B. Monk, The inverse source problem for maxwell's equations, Inverse problems 22 (3) (2006) 1023.

[41] A. Lakhal, A. Louis, Locating radiating sources for maxwell's equations using the approximate inverse, Inverse Problems 24 (4) (2008) 045020.

[42] E. A. Marengo, A. J. Devaney, R. W. Ziolkowski, Inverse source problem and minimum-energy sources, JOSA A 17 (1) (2000) 34–45.

[43] A. J. Devaney, E. A. Marengo, M. Li, Inverse source problem in nonhomogeneous background media, SIAM Journal on Applied Mathematics 67 (5) (2007) 1353–1378.

[44] G. Gragnani, M. D. Mendez, Improved electromagnetic inverse scattering procedure using non-radiating sources and scattering support reconstruction, IET microwaves, antennas & propagation 5 (15) (2011) 1822–1829.

[45] J. Sylvester, Notions of support for far fields, Inverse Problems 22 (4) (2006) 1273.

[46] A. Beck, M. Teboulle, A fast iterative shrinkage-thresholding algorithm for linear inverse problems, SIAM journal on imaging sciences 2 (1) (2009) 183–202.

[47] S. C. Park, M. K. Park, M. G. Kang, Super-resolution image reconstruction: a technical overview, Signal Processing Magazine, IEEE 20 (3) (2003) 21–36.

[48] I. Daubechies, M. Defrise, C. De Mol, An iterative thresholding algorithm for linear inverse problems with a sparsity constraint, arXiv preprint math/0307152.

[49] A. Abubakar, P. Van den Berg, S. Semenov, Two-and three-dimensional algorithms for microwave imaging and inverse scattering, Journal of Electromagnetic Waves and Applications 17 (2) (2003) 209–231.

[50] T. Rubk, P. M. Meaney, P. Meincke, K. D. Paulsen, Nonlinear microwave imaging for breast-cancer screening using gauss–newton's method and the cgls inversion algorithm, Antennas and Propagation, IEEE Transactions on 55 (8) (2007) 2320–2331.

[51] P. M. Meaney, M. W. Fanning, D. Li, S. P. Poplack, K. D. Paulsen, A clinical prototype for active microwave imaging of the breast, Microwave Theory and Techniques, IEEE Transactions on 48 (11) (2000) 1841–1853.

[52] J. De Zaeytijd, A. Franchois, C. Eyraud, J.-M. Geffrin, Full-wave three-dimensional microwave imaging with a regularized gauss–newton method—theory and experiment, Antennas and Propagation, IEEE Transactions on 55 (11) (2007) 3279–3292.

[53] J. De Zaeytijd, A. Franchois, Three-dimensional quantitative microwave imaging from measured data with multiplicative smoothing and value picking regularization, Inverse Problems 25 (2) (2009) 024004.

[54] K. Cools, F. P. Andriulli, E. Michielssen, A Calderón multiplicative preconditioner for the PMCHWT integral equation, IEEE Transactions on Antennas and Propagation 59 (12) (2011) 4579–4587.

[55] O. D. Kellogg, et al., Foundations of potential theory, New York., 1929.

[56] J. L. A. Quijano, G. Vecchi, Field and source equivalence in source reconstruction on 3D surfaces, Progress In Electromagnetics Research 103 (2010) 67–100.

[57] S. M. Rao, D. Wilton, A. W. Glisson, Electromagnetic scattering by surfaces of arbitrary shape, IEEE Transactions on Antennas and Propagation 30 (3) (1982) 409–418.

[58] D. H. Schaubert, D. R. Wilton, A. W. Glisson, A tetrahedral modeling method for electromagnetic scattering by arbitrarily shaped inhomogeneous dielectric bodies, IEEE Transactions on Antennas and Propagation 32 (1) (1984) 77–85.

[59] W. C. Chew, J.-M. Jin, E. Michielssen, J. Song, Fast and Efficient Algorithms in Computaional Electromagnetics, ARTECH house, 2001.

[60] M. Leibfritz, F. Landstorfer, T. Eibert, An equivalent source method to determine complex excitation levels of antenna arrays from near-field measurements, in: Antennas and Propagation, 2007. EuCAP 2007. The Second European Conference on, IET, 2007, pp. 1–7.

[61] F. Andriulli, A. Tabacco, G. Vecchi, Solving the EFIE at low frequencies with a conditioning that grows only logarithmically with the number of unknowns 58 (5) (2010) 1614 –1624. `doi:10.1109/TAP.2010.2044325`.

[62] F. Andriulli, Loop-star and loop-tree decompositions: analysis and efficient algorithms 60 (5) (2012) 2347 –2356. `doi:10.1109/TAP.2012.2189723`.

[63] F. Andriulli, K. Cools, I. Bogaert, E. Michielssen, On a well-conditioned electric field integral operator for multiply connected geometries, IEEE Transactions on Antennas and Propagation 61 (4) (2013) 2077–2087.

[64] F. P. Andriulli, K. Cools, H. Bagci, F. Olyslager, A. Buffa, S. Christiansen, E. Michielssen, A multiplicative calderon preconditioner for the electric field integral equation 56 (8) (2008) 2398–2412.

[65] J.-C. Nedélec, Acoustic and Electromagnetic Equations, Springer, 2000.

[66] A. Buffa, S. H. Christiansen, A dual finite element complex on the barycentric refinement, Math. Comp. 76 (260) (2007) 1743–1769.